\newcommand{\be}{\begin{eqnarray}}
\newcommand{\ee}{\end{eqnarray}}

\newcommand{\lsim}{\lesssim}

\documentclass[12pt]{article}

\usepackage{amsmath,amssymb}
\usepackage{graphicx}
\usepackage{cite}
\usepackage{bm}

\def\theequation{\arabic{section}.\arabic{equation}}
\newcommand{\eq}[1]{(\ref{#1})}

\begin{document}

\renewcommand{\thefootnote}{\fnsymbol{footnote}}
\renewcommand{\vec}[1]{{\bf #1}}

\vskip 15mm

\begin{center}

{\Large Comments on Thermodynamics of Supersymmetric Matrix Models}

\vskip 4ex

A.V. \textsc{Smilga}\,$^{1}$,

\vskip 3ex

$^{1}\,$\textit{SUBATECH, Universit\'e de
Nantes,  4 rue Alfred Kastler, BP 20722, Nantes  44307, France
\footnote{On leave of absence from ITEP, Moscow, Russia.}}
\\
\texttt{smilga@subatech.in2p3.fr}
\end{center}

\vskip 5ex

\begin{abstract}
\noindent We present arguments that the structure of the spectrum of the 
supersymmetric matrix model with 16 real supercharges in the large $N$ limit
is rather  nontrivial, involving besides the natural energy scale 
$\sim \lambda^{1/3} = (g^2 N)^{1/3}$ also a lower scale $\sim \lambda^{1/3} 
N^{-5/9}$.
This allows one to understand a nontrivial behaviour of the mean internal energy
of the system
$ E   \propto T^{14/5} $
predicted by AdS duality arguments.
 \end{abstract}

\renewcommand{\thefootnote}{\arabic{footnote}}
\setcounter{footnote}0
\setcounter{page}{1}

\section{Introduction}

The AdS/CFT duality is an efficient method allowing one to obtain nontrivial predictions
for many observables in certain supersymmetric field theories at strong coupling \cite{duality}. 
Most results were derived for ${\cal N}=4 \ \ 4D$ supersymmetric 
$SU(N)$ Yang--Mills theory in the 't Hooft limit $N \to \infty$ with fixed and large
$\lambda = g^2 N$. The wonderful Maldacena conjecture that the properties of this theory
at $\lambda \gg 1$ can be derived by studying classical solutions of $10D$ supergravity
is not proven now. However, it was verified in several nontrivial cases where {\it exact}
solution is known. Arguably, the most lucid example is  the circular
supersymmetric Wilson loop \cite{loop}. For large $N$, its  vacuum average  can be perturbatively evaluated in any order in 
$\lambda$. The sum of the perturbative series is  
 \be
\label{WCexact}
\left< W \right>_{\rm circle} \ =\ \frac {2I_1(\sqrt{\lambda})}{\sqrt{\lambda}} \ .
 \ee
On the other hand, the same quantity can be calculated at large $\lambda$ on the AdS side. The result
   \be
\label{WCAdS}
 \left< W \right>_{\rm circle} \ =\  \sqrt{\frac 2\pi } e^{\sqrt{\lambda}} 
 \frac 1 {\lambda^{3/4}}\left[1 - \frac 3 {8\sqrt{\lambda}}  + \ldots \right]\ .
 \ee
coincides exactly with the asymptotics and preasymptotics of \eq{WCexact}. Another important example is the so called
cusp anomalous dimension \cite{cusp}.

Besides vacuum averages and the scattering amplitudes, one can also
calculate thermodynamic characteristics. Thus, 
 the mean energy density of ${\cal N} = 4\ \ \  4D$ SYM system at nonzero temperature was evaluated at strong coupling at leading
\cite{Kleblead} and subleading \cite{Klebsub} order.
The result is 
 \be
\label{EKleb}
  E  =   \frac {\pi^2 N^2}{2} T^4 \left[\frac 34 + \frac {45}{32} \frac {\zeta(3)}{\lambda^{3/2}} + \ldots  \right]
 \ee
(the coefficient in front of $T^4$ is the coefficient in the Stefan-Boltzmann law).
In this case the exact result for the function $f(\lambda)$ multiplying the factor $\pi^2 N^2 T^4/2$ in the expression for $E$ is not known 
(though asymptotic expansion
\eq{EKleb} matches perfectly the known perturbative expansion of $f(\lambda)$ 
at small $\lambda$), and it is  not thus clear how the 
nontrivial coefficient $3/4$ in the strong coupling asymptotics 
is obtained, if staying in the framework of field theory and not going to the
AdS side. 

Duality relationships can be established and duality predictions can be made, however, not only for 
$4D$ theory, but also for its low-dimensional ``sisters'' obtained by dimensional reduction. In particular, by studying
a certain charged black hole solution in $10D$ supergravity, one can evaluate the average internal energy of the 
supersymmetric quantum mechanical system obtained from ${\cal N} = 4 \ \ \ 4D$ SYM by keeping there only zero spatial field
harmonics.
  The model involves 8 complex or 16 real supercharges.
The ${\cal N} = 4 \ \ \ 4D$ SYM model can in turn be obtained by dimensional reduction from ${\cal N}= 1$ 10-dimensional theory.
To distinguish it from the models obtained by dimensional reduction from $4D$ and $6D$ ${\cal N} = 1$ theories, 
we will refer to it as ``$10D$ SQM model''.  
Its Hamiltonian is
 \be
\label{HamSQM} 
H \ =\  \frac 12 E^a_i E^a_i + \frac {g^2}4 f^{abe} f^{cde} A^a_i A^b_j A^c_i A^d_j + \frac {ig}2 f^{abc} \lambda_\alpha^a
(\Gamma_i)_{\alpha\beta} \lambda^b_\beta A^c_i
\ ,
 \ee
where $i,j = 1,\ldots, 9, \ a = 1, \ldots , N^2-1$ , and $\alpha, \beta = 1, \ldots, 16$. $E^a_i$ are canonical momenta
for the bosonic dynamic variables $A^a_i$. Now, $\lambda_\alpha^a$ are Majorana fermion variables lying in the {\bf 16}- plets of
$SO(9)$. $\Gamma_i$ are 9--dimensional (real and symmetric) $\Gamma$-matrices. 
One can introduce $8(N^2-1)$ holomorphic fermion variables, 
 \be
\label{mu}
\mu^a_1 = \lambda_1^a + i\lambda_9^a\,, \ \ \ \ldots, \ \ \ \mu^a_8 = 
\lambda_8^a + i\lambda_{16}^a, \nonumber \\ 
 \bar{\mu^{a1}} = \lambda_1^a - i\lambda_9^a\,,   \ \ \ \ldots, \ \ \ \bar{\mu^{a8}}
 = \lambda_8^a - i\lambda_{16}^a\ ,
\ee
 such that 
the wave functions depend on $A^a_i$ and on $\mu^a_{1,\ldots,8}$, while   
$\bar\mu^{a |{1,\ldots,8}}$ are the fermion 
canonical momenta, $\bar\mu = \partial/\partial \mu$. 
Only the states $\Psi$ satisfying the Gauss law constraint
 \be
\label{Gauss}
\hat{G}^a \Psi  \ =\ f^{abc}\left( A^b_i \hat{E}^c_i - \frac i2 \lambda_\alpha^b \lambda_\alpha^c \right) \Psi = 0\ . 
 \ee
should be kept in the spectrum.

In $6D$ and $4D$ theories, holomorphic fermion variables
are defined more naturally as Weyl fermions lying in the complex representations of the rotational groups $SO(3)$ and $SO(5)$ 
(the spinor representation is real in  $SO(9)$ ). For example, in
$4D$ theory, the third term in the Hamiltonian is 
 \be
\label{ferm46D}
-ig f^{abc} A_i^a \bar \lambda^{b\alpha} (\sigma_i)_\alpha^{\ \beta}  \lambda_\beta^c,\ \ \ \ \ \ \   
\alpha, \beta = 1,2 
 \ee
($\lambda^{a\alpha} \equiv \partial/\partial \lambda^a_\alpha $).
We see that one can define in this case (and also in the $6D$ case) 
the fermion charge $F = \lambda^a_\alpha \bar \lambda^{a\alpha}$ that commutes 
with the Hamiltonian. In the $10D$ case, the fermion charge $\mu^a_\alpha \bar\mu^{a\alpha}$ is not conserved.

 The coupling constant  carries dimension here, $[g^2] = m^3$. The natural energy scale of the theory is thus
 \be
\label{Echar}
E_{\rm char} \sim (g^2 N)^{1/3} \equiv \lambda^{1/3}\ .
\ee

The duality prediction for the average internal energy is 
\cite{Kabat,japsublead} 
\footnote {We sketch the derivation of the leading asymptotics $\propto T^{14/5}$ in the Appendix.}
 \be
\label{145}
 \left< \frac E{N^2}  \right>_{T \ll \lambda^{1/3}} \ \approx \ 
7.41 \lambda^{1/3} \left( \frac T {\lambda^{1/3}} \right)^{14/5} 
\left[1 + O\left( \frac T {\lambda^{1/3}} \right)^{9/5} \right] \ .
  \ee

A question arises whether this rather nontrivial critical behaviour can be understood in terms of the dynamics of
the system \eq{HamSQM} without going to the supergravity side. Even though the system \eq{HamSQM} is complicated, 
it is just a QM system 
with large, but finite (for finite $N$) number of degrees of freedom. The analysis of its dynamics at strong coupling is
{\it a priori} a much more simple task than the analysis of a strongly coupled field theory.
And, indeed, in recent papers \cite{japsublead,japlead} 
(see also \cite{CatWis}) 
a numerical analysis of the system \eq{HamSQM} was performed. The results are in
a good agreement with \eq{145}. Can one understand it also analytically (staying firmly on the SQM side) ?
Our answer to this question is positive.

However, before giving this answer (it will be presented by the end of the next section), we are in a position to describe a proper 
context where the question should be posed and remind
some well-known facts.

\section{Thermodynamics and the spectrum} 
\setcounter{equation}0

As a warm-up, consider the harmonic oscillator, $H = (p^2 + \omega^2 x^2)/2$, at finite temperature. The partition
function is
 \be
\label{Zosc} 
Z \ =\ \sum_{n=0}^\infty \exp\left\{-\beta \omega \left( n + \frac 12 \right) \right \} 
 \ =\ \frac 1{2 \, {\rm sinh} \frac {\beta \omega} 2} \ .
 \ee
At large temperatures $T = \beta^{-1}$,
 \be
\label{ZhighT}
 Z_{T \gg \omega} \ \approx \ \frac T\omega \ .
 \ee 
The latter result can also be obtained semiclassically
 \be
\label{Zoscsemicl}
Z_{\rm high\ T} \ \approx Z_{\rm semicl} = \int \frac {dp dx}{2\pi} \exp\left\{ -\frac \beta 2 (p^2 + \omega^2 x^2) \right\} \ =\ \frac T\omega
\ .
 \ee
The mean energy is
 \be
\label{Eosc}
\langle E \rangle_T \ =\ -\frac \partial {\partial \beta} \ln Z = \frac \omega{2 \, {\rm tanh} \frac {\beta\omega}2 }\ .
 \ee 
At low temperatures, $\langle E \rangle_T \approx \omega/2 + O(e^{-\beta \omega})$, while at high temperatures,
 \be
\label{EhighT}
 \langle E \rangle_{T \gg \omega} \ \approx \  T \ .
 \ee 

The behaviour $ \langle E \rangle_{\rm high \ T} \ \propto \  T$ is characteristic not only for the oscillator, but 
for {\it any} reasonable QM system. Basically, it is the analog of the Stefan-Boltzmann law in zero spatial dimensions.
 For the oscillator
 with several (many) degrees of freedom $\#$, this number multiplies $T$ in  the high--temperature estimate for 
$\langle E \rangle_T$. 

\subsection{YM matrix models.}

For purely bosonic matrix models with the Hamiltonian being the sum of two first terms in \eq{HamSQM}, 
the pattern of the spectrum and the temperature dependence of the average energy and other thermodynamic functions
 is clearly understood and is  
much simpler than that for supersymmetric models. It does not depend much on whether the model is obtained by reduction from $4D$
YM theory ($i = 1,2,3$), $6D$ theory ($i = 1,\ldots,5$), or $10D$ theory   ($i = 1,\ldots,9$).

Let us first estimate the energy of the ground state. The simplest variational Ansatz is 
 \be
\label{Psivar}
\Psi_0 \propto \exp\left\{ - \alpha (A^a_i)^2 \right\} \equiv e^{-\alpha A^2} \ .
 \ee
The contribution of the kinetic term in the Hamiltonian to $E_{\rm var}$ is estimated as
 \be
\label{Evarkin}
E_{\rm var}^{\rm kin} \ \approx  \alpha N^2 \ ,
 \ee
where we have kept only the dependence on $N \gg 1$, not worrying about the dependence on $D$ and about
numerical coefficients. (Note that $\langle A^2 \rangle_0 \sim N^2/\alpha$ in this limit.)
 To estimate the contribution of the potential part, use
 $$ \langle A^a A^b \rangle \sim \frac {\langle A^2 \rangle_0 \delta^{ab}} {N^2}, \ \ \ \ \ \ 
\langle A^a A^b A^c A^d \rangle  \sim  \frac {\langle A^4 \rangle_0 \left( \delta^{ab} \delta^{cd} + \delta^{ac} \delta^{bd} 
+ \delta^{ad} \delta^{bc} \right)}{N^4}   $$
(irrelevant spatial indices are suppressed). 
We obtain
  \be
\label{Evarpot}
E_{\rm var}^{\rm pot} \ \approx \frac {g^2\langle A^4 \rangle_0}N \sim \frac {g^2 N^3}{\alpha^2}   \ .
 \ee
Adding this to \eq{Evarkin} and minimizing over $\alpha$, we find
 \be
\label{varest}
\alpha \sim \lambda^{1/3}, \ \ \ \ \ \ \ \ \ \ \ E_0 \sim N^2 \lambda^{1/3} \ .
 \ee
In other words, the estimate for the vacuum energy is obtained by multiplying the natural
energy scale \eq{Echar} by the number of degrees of freedom $\sim N^2$. The characteristic size of the wave function (`` extent of space''
 in
the terminology of Refs.\cite{japsublead,japlead}) is also
 determined by this scale,
 \be
\label{extent}
{\rm extent \ of \ space}  
\sim  \frac {\langle A^2 \rangle_0}{N^2} \sim \frac 1\alpha \sim \frac 1{E_{\rm char}} \ . 
 \ee
The gap between the first excited state and the vacuum state is also of order $E_{\rm char}$.  

The partition function at high temperatures $T \gg \lambda^{1/3}$ is easily evaluated by semiclassical methods \cite{jaCMP}.
We have
 \be
\label{Zbossemicl}
Z^{\rm high\ T} \approx \int \prod_{ai}  \frac {dE^a_i dA^a_i }{2\pi}  \prod_a \delta(G^a)e^{-\beta H}  = \nonumber \\
\int \prod_{ai}  \frac {dE^a_i dA^a_i }{2\pi} e^{-\beta H}  \prod_a \frac {dA_0^a}{2\pi}  \exp\{iA^a_0 f^{abc} A^b_i E^c_i \} \ . 
 \ee
The integral is saturated by the characteristic values 
 $$(A^a_i)_{\rm char} \sim T^{1/4} \lambda^{-1/4}, \ \ \ \ (E^a_i)_{\rm char} \sim \sqrt{T}, \ \ \ \ \ \ 
(A^a_0)_{\rm char} \sim (A^a_i)_{\rm char}^{-1} (E^a_i)_{\rm char}^{-1} \sim \lambda^{1/4} T^{-3/4}  $$
and is estimated as 
\be
\label{Zdiscr}
Z_{\rm semicl} \sim \left( \frac T{\lambda^{1/3}} \right)^{\frac 34 N^2 (D-2)} \ ,
 \ee
which gives \cite{japhighT}
 \be
\label{Ediscr}
\langle E\rangle^{\rm YM}_{\rm high \ T} \ \sim \ \frac 34 N^2 (D-2) T \ .
 \ee   
One can also evaluate corrections to this leading order semiclassical result. To estimate the next-to-leading correction,
 one should roughly speaking insert the factor $\sim \beta^2 \partial^2 V/ (\partial A^a_i)^2  \sim \beta^2 \lambda 
(A^a_i)^2$ in the
integrand in \eq{Zbossemicl} ($V$ is the potential) \cite{Feynman,jaCMP}. This gives
\footnote{The parameter $c$ was evaluated numerically for $D=10$ in Ref.\cite{japhighT}. It is negative.} 
 \be
 \label{Zsemicorr}
 Z_{\rm high\ T} \sim \left[ \frac T{\lambda^{1/3}} 
\left(1 + c \frac {\lambda^{1/2}}{T^{3/2}} \right)  \right]^{\frac 34 N^2 (D-2)}
 \ee
We see that the correction is of order one at $T \sim \lambda^{1/3}$. At  temperatures much less than the characteristic spectral gap 
$E_{\rm char} \sim \lambda^{1/3}$, \ $\langle E \rangle_T$ behaves
in the same way as  the oscillator average energy \eq{Eosc} coinciding with the vacuum energy up to exponentially small corrections.
This pattern was confirmed  by numerical calculations \cite{japbos}.

\subsection{Supersymmetric models}

One important  feature distinguishing all SYM models ($10D,\ 6D$, and $4D$) from
the  YM models is the presence of continuous spectrum
associated with flat directions in the potential \cite{Wit02}. The classical potential $V \sim g^2 \sum_{ij} {\rm Tr} 
\{ [ A_i, A_j]^2 \}$ turns to zero when $ [ A_i, A_j] = 0$, i.e. when all $A_i$ belong to the Cartan subalgebra. 
In purely bosonic model,
this classical degeneracy is lifted by quantum corrections. It supersymmetric models, this does not happen. As a result, 
the states can smear along the flat directions, the motion becomes  infinite, and the continuous branch of the spectrum
exists. For low energies, $E \ll \lambda^{1/3}$,  the wave functions of these states can be evaluated  in the framework of
 the Born--Oppenheimer approximation \cite{Wit02,jaBO,supmemb,Nicolai,sKac,anygroup}. In the leading order, they can be 
chosen in the form
 \be
\label{BOAns}
\Psi_{\rm continuous} \ \approx \ \chi(x_{\rm slow}) \, \psi_{A_{\rm slow}} \left( x_{\rm fast} \right) \ . 
 \ee
$x_{\rm slow}$ in this expression stand for $A^{\tilde a}_i$ 
and their fermionic partners. 
The Cartan subalgebra index ${\tilde a}$ runs from 1 to $r = N-1$. $ x_{\rm fast}$ are the transverse to the valley components
of $A$ and their fermionic partners. The motion across the valley is described by the Hamiltonian of supersymmetric oscillator  with
the frequency $\sim g |\vec{A}|$. 
$\psi_{A_{\rm slow}}(x_{\rm fast})$ is its ground state. The function $\chi(x_{\rm slow})$ is the eigenfunction of the effective
Born-Oppenheimer Hamiltonian. To leading order, the latter is just the Laplacian 
 \be
\label{Laplac}
H_{\rm BO} \sim \ -\frac 12 \frac {\partial^2}{(\partial A^{\tilde a}_i )^2} + \ldots
 \ee
such that  $\chi \propto \exp\{ik^{\tilde a}_i A^{\tilde a}_i \}$. 

In $4D$ and $6D$ theories where the conserved fermion charge exists, one can ask what is its value  
for the continuum
spectrum states \eq{BOAns}. Consider for simplicity 
 $4D$ theory. The fast oscillator Hamiltonian depends on $(D-2)N(N-1) = 2N(N-1)$ real
 bosonic variables and 
$2N(N-1)$ holomorphic fermion variables [ $N(N-1)$ being the total number of roots in $SU(N)$]. 
When $N=2$,  the ground state of the fast Hamiltonian has the structure \cite{supmemb}
  \be
 \label{fastN2} 
 \psi_C(x_{\rm fast}) \sim \exp \left\{ - \frac {gC}2 (A^a_m)^2 \right\} 
\left\{ \lambda^{b\alpha} \lambda_\alpha^b + i \epsilon^{bc}
\lambda^{b\alpha} (\sigma_3)_\alpha^{\ \beta} \lambda^c_\beta \right\} \ ,
 \ee
where we have directed the slow variable $A^3_i$ along the 3-d spatial axis such that $A^3_i = C\delta_{3i}$. Here 
the indices $a,b = 1,2$ are transverse color indices  and the index $m = 1,2$ is the transverse spatial index;
$\lambda^{a\alpha} = \epsilon^{\alpha\beta} \lambda^a_\beta$. 
The fermion charge of the function \eq{fastN2} is 2. For larger $N$, the fast ground state wave function 
involves $N(N-1)/2$ such fermion factors, each factor carrying the charge 2. 
The total fermion charge of $\psi_{A_{\rm slow}} \left( x_{\rm fast} \right)$ is thus $F_{\rm fast}(N) = N(N-1)$. Speaking of
$\chi(x_{\rm slow})$, it may carry fermion charges from $F_{\rm slow} = 0$ to $F_{\rm slow} = 2(N-1)$. 
All together we have $2^{2(N-1)}$ families
of continuum spectrum states carrying fermion charges from $N(N-1)$ to $(N+2)(N-1)$. Their average  charge is $N^2-1$, the half 
of the maximal  fermionic charge $F_{\rm max}$.

The partition function for the system with continuous spectrum is infinite. For example, for the Hamiltonian
$H = p^2/2$,
 \be
\label{Zp2}
 Z = \int \frac {dpdx}{2\pi} e^{-\beta p^2/2} \ =\ L \sqrt { \frac T{2\pi}} \ ,
 \ee 
with $L \to \infty$. However, the average energy $\langle E \rangle_T$ defined as in \eq{Eosc} does not 
depend on the infinite factor $L$ 
and is equal to $T/2$. The Hamiltonian \eq{Laplac} involves $(D-1)(N-1)$ degrees of freedom and we thus  obtain
 \be
\label{Zcont}
Z \sim \left(  L \sqrt { \frac T{2\pi}} \right)^ {(D-1)(N-1)} \ \sim
\ \left( \frac T\mu \right)^{(D-1)(N-1)/2}
 \ee
(where an infrared regulator $\mu$ carrying dimension of energy is introduced) 
and 
  \be
 \label{Econt}
 \langle E \rangle_T  \approx \frac {(D-1)(N-1)T}2 \ .
 \ee
This result has nothing to do with the supergravity prediction \eq{145} ! 

The estimate \eq{Zcont} for the partition function contradicts, however, path integral estimate.
The latter is definitely correct (and hence the former is definitely wrong) at high temperatures, 
$T \gg \lambda^{1/3}$, when fermions and higher Matsubara modes decouple and the partition function is given
by the semiclassical estimate \eq{Zdiscr}, the same as for the pure YM system [the fermions could only affect the 
coefficient $c$ in \eq{Zsemicorr} ].

The paradox is resolved by noting that the spectrum of our system involves besides the continuous spectrum also the 
discrete spectrum with
normalized states. Consider first $4D$ theory. In this case, the normalized discrete spectrum states of  pure bosonic theory
represent  also eigenstates of the 
 full Hamiltonian: the fermion term \eq{ferm46D} gives zero when acting on the states of zero fermion charge.
These normalized eigenstates have the energy $\sim N^2 \lambda^{1/3}$ as in \eq{varest}. The characteristic gap between
the lowest and excited normalized states in the sector $F=0$ is of order $\lambda^{1/3}$.   
 Acting on these states
by supercharges $Q_\alpha$, we can obtain normalized eigenstates of the full Hamiltonian in the sectors $F=1$ and $F=2$. 
By the same token, one can construct  the states in the sectors $F = F_{\rm max} = 2(N^2-1)$, $F =  F_{\rm max} -1$ and $F =  F_{\rm max} -2$. 
 Little is known about the structure
of normalized eigenstates in the sectors with other values of $F$. It is natural to suggest, however, that they
also  exist  and that some of these states (probably, in the sectors with $F \sim\ F_{\rm max}/2$ ) 
may have energy as low as  $\lambda^{1/3}$ (without the $N^2$ factor).  

If the continuum states did not exist, the partition function at high temperatures would be given by the estimate 
\eq{Zdiscr} (with $D=4$) and $\langle E \rangle_T$ by the estimate \eq{Ediscr}. One can observe now that the latter
is much larger than \eq{Econt} at large $N$. Heuristically, this means that at large $N$ the average energy is determined
by the discrete spectrum states, while the continuum states are irrelevant. 

Thinking a little bit more in this
direction, one could judge that this heuristic impression is wrong because one cannot just add the estimates 
\eq{Ediscr} and \eq{Econt} and observe that \eq{Ediscr} dominates. One should add the contributions to the partition function
rather than to the energy. The continuum spectrum contribution to the partition function \eq{Zcont} involves an
infinite factor $\mu^{-(D-1)(N-1)/2}$ and always dominates.

Thinking still more, one finds, however, that the limits $\mu \to 0$ and $N \to \infty$ do not commute. 
 \begin{itemize}
\item  At finite $N$ and small enough $\mu$, the continuum contribution \eq{Zcont} to the partition function 
dominates and one can forget
about discrete spectrum.
 \item
At finite $\mu$ and large enough $N$,  the discrete spectrum contribution \eq{Zdiscr} to the 
partition function dominates and one can forget
about continuum. For large enough temperatures, $T \gg \lambda^{1/3}$, 
one can also forget
about continuum at small $N$ down to $N=2$.   
 \end{itemize}

Thus, introducing the infrared regulator and playing with this parameter, one can get rid of the contribution \eq{Econt}. If
 $N$ is large enough, one should be able to do it  for {\it any} temperature and not necessary for high
 temperatures $T > \lambda^{1/3}$ where the estimate \eq{Ediscr} is derived. In numerical calculations 
\cite{japlead,japsublead},
no infrared regulator was introduced, but the algorithm was chosen such that the continuum spectrum effects were effectively 
suppressed.
The functional integral for $Z$ was done by Metropolis algorithm with initial values of all components 
$A_i^a$ chosen to be of the same order and not very large. It was then observed that, for small $N$, this configuration 
is unstable
such that the field variables tend to smear along the flat directions. However, for larger $N$, the system penetrates the valley
only after a considerable number of iterations. The larger is $N$ and/or $T$, the more stable is the system. An effective barrier is erected.
For large $N$, one can thus evaluate
the averages {\it before} the system penetrates through this barrier and escapes along the valley. 

Unfortunately, no numerical calculations for the $6D$ system have been done yet, while existing
calculations for the $4D$ systems \cite{num4D} are not good enough to conclude about the behaviour of 
$\langle E \rangle_T$ at low $T$ and  large $N$.  The authors of
\cite{japlead,japsublead} concentrated on studying the  $10D$ system, where they were able to compare their
results with the supergravity predictions. It seems to us very important to perform the measurements with large enough $N$ 
and good enough statistics also for $4D$ and $6D$ systems
and {\it compare} the results with those obtained in the $10D$ case. 

What predictions can be made for the behaviour
of $\langle E \rangle_T$ in the $4D$ and $6D$ cases ? 
Let us {\it assume}
\footnote{We will rediscuss this assumption in Sect. 3.}
 the pattern of discrete spectrum states spelled out above: the lowest such state has the energy
$\sim \lambda^{1/3}$ and higher excited states behave roughly in the same way as in the
 purely bosonic matrix models up to an overall shift
\be
\label{shift}
E_{\rm vac}^{\rm YM} \sim N^2 \lambda^{1/3} \ \longrightarrow \ E_{\rm vac}^{\rm SYM} \sim 0\ .
 \ee
It follows  then  that, 
in the limit $N \to \infty$, with the continuum spectrum effects filtered out as explained above, 
$\langle E \rangle_T$ behaves in the same way as for purely bosonic models, i.e. is given by the estimate \eq{Ediscr}
at $T > \lambda^{1/3}$ and approaches zero exponentially fast at $T \to 0$. When $N$ is large, but  finite, this behaviour
is valid down to $T_\star \sim \lambda^{1/3}/\ln N$ such that  $\langle E \rangle_{T_\star} \sim NT_{\star}$ and coincides
with the continuum estimate \eq{Econt}. At still lower temperatures, the system is not contained in the region around
$A \sim 0$, but penetrates the barrier and is smeared along the valley. The estimate \eq{Econt} for $\langle E \rangle_T$
is valid in this region.    

We are prepared now to go over to $10D$ theory and make finally an original remark that is {\it raison d'\^etre} for 
this paper. $10D$ theory  has one important feature that is lacking in $4D$ and $6D$ theories: on top of continuum
low-energy states and excited discrete spectrum states, it involves a normalized
{\it vacuum} state with zero energy. Its existence was first discovered in \cite{Yiidr} from  calculations of integrals
for Witten index, but a clearer way to see it consists in deforming  theory by endowing
the scalar fields (in the reduced SQM model, they correspond to the components of $A_{4,\ldots,9}$ with) with mass $M$ 
\cite{PorRos}. For nonzero mass, the classical potential has  isolated zero energy minima
at large values of the fields. Generically, there are several (many) such minima \cite{sKac}, but for
the $SU(N)$ groups, there is only one classical vacuum. If mass is large, the walls of the potential well around
this classical vacuum are steep and the wave function of the corresponding  quantum vacuum is a localized oscillator
wave function. By continuity, a localized wave function exists also for small values of mass. It is a natural hypothesis that
the vacuum state remains normalizable also in the limit $M \to 0$.

Another argument comes from Born-Oppenheimer analysis of the vacuum wave function in the valley. {\it If} the normalized
state with zero energy exists, its wave function in the valley should be represented in the form
\eq{BOAns} with $\chi(x_{\rm slow})$ representing the eigenfunction of the effective Hamiltonian \eq{Laplac} with zero eigenvalue.
The full wave function should be annihilated by supercharges $Q_\alpha$ and that means that
   $\chi(x_{\rm slow})$ should be annihilated by effective supercharge acting in  Hilbert space of slow variables. 
One can show that normalized solutions to the equation $Q^{\rm eff}_\alpha \chi(x_{\rm slow}) \ =\ 0$,
 supplemented 
by the requirement of Weyl invariance of $  \chi(x_{\rm slow})$ following from gauge invariance of the full wave function, 
do {\it not} exist in $4D$ and $6D$ theories, but the solution {\it exists} in the $10D$ case. It was explicitly constructed
for $SU(2)$ \cite{asN2} (see also \cite{sKac} for pedagogical explanations) and for $SU(3)$ \cite{asN3}. In the simplest
$N=2$ case, the asymptotic vacuum wave function has the form 
   \be
\label{aschi}
\chi_{\rm vac}(A_i,\mu_{1,\ldots,8}) \propto ({\bf 44}^{\rm ferm})_{ij}
\partial_i \partial_j \frac 1 {|{\bf A}|^7} \ ,
 \ee
where $\vec{A} \equiv \vec{A}^3$ and $({\bf 44}^{\rm ferm})_{ij}$ is a fermionic structure representing 
the ${\bf 44}$-plet of $SO(9)$. 

 The result \eq{aschi} is obtained in the leading Born--Oppenheimer order. The corrections to this result are small
when the corrections to the effective Hamiltonian are small. First subleading corrections to the effective Hamiltonian
 are known. Supersymmetry prevents the generation of the potential on the valley. In $10D$ theory, there are also no 
corrections to the metric, i.e. to the term $\propto E^2$ in the Hamiltonian. 
\footnote{This result obtained first in \cite{Stern} has much in common with $4D$ nonrenormalization theorems \cite{SAV}.}
However, there are corrections $\propto E^4$. The exact form of these corrections in the $N=2$ case is \cite{Becker}
 \be
\label{2corr}
H_{\rm eff}  = \frac {\vec{E}^2}2 + \frac {15}{16} \frac {|\vec{E}|^4}{g^3|\vec{A}|^7} + \ldots 
+ {\rm terms\ with\ fermions}.
 \ee
The correction is small iff $g|\vec{A}|^3 \gg 1$ (we used $\vec{E}^2 \sim 1/\vec{A}^2$). 
This is also the region 
where the expression 
\eq{aschi} for the asymptotic vacuum wave function is valid. On the other hand, when  $g|\vec{A}|^3 \lsim 1$, the separation
of slow and fast variables does not work, and the wave function depends on all components of $A^a_i$ in a complicated way.
Thus, the characteristic size of the vacuum wave function is of order $A_{\rm char}^2 \sim g^{-2/3}$, which is the same as
\eq{extent}, if disregarding $N$--dependence there.

For $N >3$, the asymptotic wave function has not been constructed explicitly. We can estimate, however, its characteristic
size as such $A_{\rm char}$ that the corrections $\sim E^4/A^7$ in the effective Hamiltonian are of the same order as the 
leading term. To this order, the effective Hamiltonian is known for any $N$ \cite{Okawa},
  \be
\label{Ncorr}
H_{\rm eff}(N)   = \sum_{n =1}^N |\vec{E}^n|^2 + \frac {15}{16} 
\sum_{n > m}^N 
\frac {|\vec{E}^n - \vec{E}^m|^4}{g^3|\vec{A}^n - \vec{A}^m|^7} + \ldots ,
 \ee
where we assumed $ \vec{\hat A}  =  {\rm diag} (\vec{A}^1, \ldots, \vec{A}^N)$ and
 $ \vec{\hat E}  =  {\rm diag} (\vec{E}^1, \ldots, \vec{E}^N)$ with $\sum_n \vec{A}^n 
 = \sum_n \vec{E}^n  = 0$. The second term in \eq{Ncorr} represents the sum over all positive
roots of $SU(N)$ (a generalization to an arbitrary group is thus trivial). For large $N$, it involves of order
$N^2$ terms, while the first term has $N$ terms. The estimate for $A_{\rm char}$ is obtained from the condition
$$  \frac {N}{A_{\rm char}^2} \  \sim  \  \frac {N^2}{g^3 A_{\rm char}^{11}}\ .  $$
 We have
 \be
\label{sizevac}
A_{\rm char}^2 \ \sim \ \frac {N^{2/9}}{g^{2/3}} \ \sim \ {N^{5/9}}{\lambda^{-1/3}}\ .
\ee
In Ref.\cite{Okawa}, also two-loop corrections to the effective Hamiltonian were evaluated. They are estimated 
as $\sim N^3 E^6/(g^6A^{14})$ and  are of the same order as the leading term at the scale \eq{sizevac}.
This is probably   as well true  for higher loop corrections, which should be of order
 $$
H^{n \ {\rm loop}} \ \sim \ \frac {E^{2(n+1)} N^{n+1}}{g^{3n} A^{7n}} \ .
 $$
We see that the estimated size of the vacuum wave function turns out to be essentially {\it larger}
than the characteristic size \eq{extent} of bosonic eigenstates. The large characteristic size \eq{sizevac}
suggests the existence of the energy scale
 \be
\label{Echarnew}
 E_{\rm char}^{\rm new} \ \sim N^{-5/9} \lambda^{1/3} \ ,
  \ee
which is considerably {\it smaller} than the principal energy scale \eq{Echar}. It is natural then to
assume that the characteristic gap in the spectrum of excited states in $10D$ SYM quantum mechanics
is not $\sim \lambda^{1/3}$ as it was in purely bosonic theory, but smaller, being given by the estimate
\eq{Echarnew}. The presence of a large number of discrete spectrum states with the energies lying in the
interval $\lambda^{1/3} N^{-5/9} < E < \lambda^{1/3}$ modifies essentially the behaviour of the partition
function. It need not now be exponentially small at $T \ll \lambda^{1/3}$, but can well display a power
behaviour. Assume that
$$\langle E \rangle_T \ \sim N^2 \lambda^{1/3} \left( \frac T{\lambda^{1/3}} \right)^\gamma $$
and determine $\gamma$ from the condition $ \langle E \rangle_T \sim \langle E_{\rm cont} \rangle_T 
 \sim NT$ at $T \sim \lambda^{1/3} N^{-5/9}$. We obtain then $\gamma = 14/5$ in a remarkable agreement with the
duality prediction \eq{145}!

 The law \eq{145} implies the behaviour
 \be
\label{Zcrit}
Z(T) \ \sim \ \exp\left\{ N^2 \left( \frac T{\lambda^{1/3}} \right)^{9/5} - N \right\}
 \ee
for the partition function. The normalization (which does not affect $\langle E \rangle_T$) was chosen such
that $Z(T \sim   E_{\rm char}^{\rm new}) \sim 1$. The latter  must be true for the discrete spectrum contribution, and
the continuum spectrum contribution can match it at the scale $T \sim   E_{\rm char}^{\rm new}$ if choosing $\mu$
 of the same order. When $T >  E_{\rm char}^{\rm new}$, the discrete spectrum contribution dominates. Expressing
$Z(T)$ into the density of states $\rho(E)$, 
 \be
Z(T) = \int \rho(E) e^{-E/T} dE\ ,
 \ee
 we see that the behaviour \eq{Zcrit} implies
 \be
\label{rhoE}
\rho(E) \propto \exp\left\{ N^2 \left( \frac E{\lambda^{1/3}} \right)^{9/5} \right\}\ .
 \ee
The critical behaviour \eq{145}, \eq{Zcrit}, \eq{rhoE} should be characteristic in the intermediate region
 \be
\label{range}
\lambda^{1/3} N^{-5/9} < T,E < \lambda^{1/3} \ .
 \ee
 At larger temperatures, the laws \eq{Zdiscr}, \eq{Ediscr}
 should take over.

\section{Discussion}
\setcounter{equation}0

Let us summarize our arguments. First, we remark that $10D$ theory involves besides continuum spectrum that is 
characteristic also for $4D$ and $6D$ theories, a normalized vacuum state. We estimate a characteristic size of this state
by requiring that the loop corrections to the effective Hamiltonian at this scale are of the same order as the leading term.
This gives us a new energy scale \eq{Echarnew}, which is lower than the principal energy scale \eq{Echar}. Then we conjecture
that, on top of the vacuum state, a large family of excited states associated with this scale exists. This explains
 the critical behaviour \eq{145}. 

 The existence of new scale should also show up in other quantities. For example, it suggests that
the ``extent of space'' \eq{extent} in $10D$ supersymmetric theory should be essentially larger than for purely bosonic 
 system. In particular, the avegage \eq{extent} should grow with $N$. The existent measurements of $\langle (A^a_i)^2/N^2 \rangle_T$
in this theory \cite{japlead} give the value that is somewhat larger than in the purely bosonic case, but no growth with 
$N$ was observed. We do not understand it in the framework of our conjecture and can only express a wish that more measurements of
this quantity at larger values of $N$ and/or lower temperatures were done. Another issue that is not clear now is the range of temperatures
where the law \eq{145} should hold. We have detected only one new scale \eq{Echarnew} and no other scales. This implies that
the law \eq{145} should be valid {\it between} these scales, i.e. in the range \eq{range}. On the other hand, considering this
problem on the supergravity side, one obtains that subleading in $N$ corrections due to string degrees of freedom become essential
at $T \sim \lambda^{1/3} N^{-10/21}$ \ \cite{japsublead}, 
which is somewhat larger than $ E_{\rm char}^{\rm new}$. It is difficult to explain the appearance
of this extra scale staying on the matrix model side.

Let us go back now to the $4D$ and $6D$ models. These models do not involve a normalised vacuum state and we conjectured
[see the discussion around Eq. \eq{shift}] that the pattern of {\it excited} normalized states is roughly 
the same there as in  purely bosonic theory. On the other hand, there are some indications of the presence of a new energy
scale also for $D= 4,6$. When $D < 10$, the corrections to the moduli space metric do not vanish. For an arbitrary gauge group, 
they have the form \cite{anygroup} 
   \be
\label{corr46}
H_{\rm eff}  \ \sim \  \frac 1{c_V} \sum_j |\vec{E}^{(j)}|^2 \left[
1 +  \frac {a_D c_V}{g|\vec{A}^{(j)}|^3}  + \ldots  \right]
+ {\rm terms\ with\ fermions}\ ,
 \ee  
where $a_4 = 3/4, a_6 = 1/2$ (and $a_{10} = 0$), $\sum_j$ is the sum over all positive roots, 
$\vec{A}^{(j)} = \alpha_j (\vec{A}^{\rm Cartan})$, $\vec{E}^{(j)} = \alpha_j (\vec{E}^{\rm Cartan})$
\footnote{For example, for $SU(N)$, $\alpha_{nm}(\vec{A}^{\rm Cartan}) = \vec{A}^n - \vec{A}^m$. },  
and $c_V$ is the adjoint Casimir eigenvalue.
For $SU(N)$ with large $N$, the corrections are of order 1 at $gA_{\rm char}^3 \sim N$, which gives
  \be
\label{Achar46}
D = 4,6 : \ \ \ \ \ A_{\rm char}^2 \sim \frac {N^{2/3}}{g^{2/3}} \sim N \lambda^{-1/3} \ .
 \ee
This might be associated with the energy scale $\sim \lambda^{1/3}/N$. {\it If} assuming that a family 
of normalised excited states with a characteristic gap $\sim \lambda^{1/3}/N$ is present there, one could deduce
that the average energy behaves as
 \be
\label{T2}
\langle E \rangle_T^{4D, 6D} \ \propto \ T^2
 \ee
in the range $\lambda^{1/3}/N < T < \lambda^{1/3}$. 

We would rather lay our own bets not on \eq{T2}, but on the scenario spelled out above - no low-energy discrete spectrum states and
$ \langle E \rangle_T$ approaching zero exponentially fast at $\lambda^{1/3}/(\ln N) < T < \lambda^{1/3}$, with continuum spectrum
dependence $\langle E \rangle_T \sim NT$ taking over at still lower temperatures.
But theoretical arguments are heuristic and uncertain
 here and only (numerical) experiment can tell us what is true.

As was mentioned, a nontrivial power behaviour of $\langle E \rangle_T$ is associated with the presence of
low-energy discrete spectrum states. In principle, one can find these states by solving Schr\"odinger equation numerically.
This calculation is, however, much more difficult than the calculation of the Euclidean path integral that determines the 
partition function 
of the system. Up to now, only the simplest $4D$ system with $N=2$ was studied \cite{Wosiek}. It would be very interesting to do it 
also for higher $N$ and find out whether a lower energy scale $\sim \lambda^{1/3}/N$ shows up there.  
 
A ``natural'' behaviour of $\langle E \rangle_T$ in quantum mechanics is $\langle E \rangle_T \propto \#_{\rm d.o.f.} T$.
In our case, the power of $T$ is different,
which may be associated with the fact, that in the infinite $N$ limit, we are dealing actually not with 
quantum mechanics, but with field theory. Indeed, it is known since \cite{Hoppe} that the Hamiltonian of the supersymmetric matrix model
coincides in the infinite $N$ limit with the supermembrane mass operator
 \be
\label{supmemb}
\lim_{N \to \infty} H_{SQM\ YM} \ = \ M^2_{\rm supermembrane} = \int d^2\sigma \left[ (P'_i)^2 + \frac 12 \{X_i, X_j\}^2 +
\ {\rm fermionic\ term} \right] \ ,
 \ee
where $P_i'$ involves only nonzero modes contribution and $\{X_i, X_j\} = \epsilon^{rs} \partial_r X_i \partial_s X_j $. 
The Hamiltonian \eq{supmemb} is invariant with respect to area-preserving diffeomorphisms (this is where gauge symmetry is transformed 
to in the limit 
$N \to \infty$). Supermembrane theory is a (2+1)-dimensional field theory. For the latter, a ``natural'' law for the area
energy density is $\propto  T^3$. For sure, this law is derived for  a conventional $SO(2,1)$ invariant field theory where, in the limit when interactions
are switched off, the spectrum represents a tower of oscillators.
The model \eq{supmemb} does not have these features. Still, one can notice that $14/5$ is numerically close to 3.
\footnote{A further numerological observation is that 5 and 14 are the fourth and the fifth term in the sequence of
{\it Catalan numbers} \cite{Ilia}}.

We would like to conclude with a general remark. 
There is a fruitful strategy: whenever you do not understand something in field theory, look at a proper QM system
where the same phenomenon occurs, analyze it, and you will get chances to improve your understanding. We think that this strategy 
applies  to Maldacena's 
duality conjecture as well. The QM system \eq{HamSQM} {\it is} complicated. Still, it is less complicated than $4D$ SYM theory 
at strong coupling.
If understanding why and how duality works in the former, we will get  chances to eventually understand it ({\it prove} it )
in field theories.

\section{Acknowledgements}
I am  indebted to E. Akhmedov, E. Ivanov, and M. Konyushikhin for illuminating discussions. Special thanks are due to M. Hanada
for very useful correspondence concerning the results and numerical procedure in Refs.\cite{japbos,japlead,japsublead}.

\section*{Appendix. Supergravity derivation of the law \eq{145}.}
\def\theequation{A.\arabic{equation}}
\setcounter{equation}0

To make the paper more self-consistent, we sketch here the derivation of \eq{145} on the supergravity side.
A reader is addressed to the original papers \cite{Kabat} and to the review \cite{dualreview} for more details.

The duality conjecture  is that quantum dynamics of different dimensionally reduced descendants of 
$10D$ $SU(N)$ SYM theory
in the large $N$ limit can be accessed by studying proper classical solutions of $10D$ IIA or IIB supergravity. 
The bosonic part of the supergravity action is
 \be
\label{SUGRA}
 S \ \propto \ \int d^{10}x \sqrt{-g} \left\{  e^{-2\phi}\left[ R  + 4(\nabla \phi)^2 \right]
- \sum_p c_p F^2_{p+2} \right\}\ ,
 \ee
where  $\phi$ is the dilaton field,  $F_{p+2}$ are field strengths of
various fundamental ($p+1$)-forms ({\it Ramond-Ramond} forms) that are present in the model, and $c_p$ are irrelevant
numerical coefficients. The action \eq{SUGRA}
admits {\it black brane} solutions,
  \be
\label{brana}
ds^2 &=& f_p^{-1/2} \left[-dt^2 \left(1 - \frac {r_0^{7-p}}{r^{7-p}} \right) + \sum_{n=1}^p dy_n^2   \right] +
f_p^{1/2}\left[ \frac {dr^2}{1 - \frac {r_0^{7-p}}{r^{7-p}}} + r^2 d\Omega^2_{8-p} \right]\ , \nonumber \\
e^{-2\phi} &\propto& f_p^{(p-3)/2} \ , \nonumber \\
 A_{p+1} &=&  {\rm irrelevant} \ ,
  \ee
where 
 $f_p$ is a harmonic function in transverse directions $(x_{p+1}, \ldots, x_9)$. The simplest choice is
\footnote{Then the metric \eq{brana} is not asymptotically flat and describes a black brane or, which is more relevant
in the context of etablishing the duality correspondence, a stack of $N$ coinciding black branes {\it in  the throat}.
By adding a constant to \eq{fp}, one could obtain an asymptotically flat black brane solution.}   
 \be
\label{fp}
f_p =  \frac {R^{7-p}}{r^{7-p}} \ .
  \ee
($r^2 = \sum_{m=p+1}^9 x_m^2$ and $R$ is a constant). $d\Omega_{8-p}^2$ is the metric on $S^{8-p}$. 
For example, if $p=3$ and $r_0 = 0$, the metric is reduced to
 \be
\label{dsAdS}
ds^2 = \left[ \frac {r^2}{R^2} (-dt^2 + dy_1^2 + dy_2^2 + dy_3^3) + \frac {R^2 dr^2}{r^2}\right] + R^2 d\Omega_5^2 \ ,
 \ee
 which is the metric on $AdS_5 \times S^5$. 

The solutions with $p=3$ are relevant when discussing physics of $(p+1) = (3+1)$ -- dimensional SYM theories.
We are interested in the dynamics of $(0+1)$ theories and in the black {\it hole} solution with $p=0$,
 \be
\label{BH}
ds^2_{\rm BH} \ =\ \frac {r^{7/2}}{R^{7/2}} \left[ -dt^2 \left( 1 - \frac {r_0^7}{r^7} \right) \right] 
+ \frac  {R^{7/2}}{r^{7/2}} \left[ \frac {dr^2}{1 - \frac {r_0^7}{r^7}} + r^2 d\Omega_8^2 \right] \ .
 \ee
Let us assume that the black hole size is much less than the characteristic curvature radius of the Universe
where it sits. This means $r_0 \ll R$. The black hole \eq{BH} has a characteristic Hawking temperature and the
Bekenstein-Hawking enthropy coinciding with the volume of its horizon in Planck units.

 To find the latter, we should simply multiply the factor
$$ \
\sqrt{-g} (r = r_0) \propto \sqrt {\left( \frac {r_0^2}{r_0^{7/2}} \right)^8} \ =\ \frac 1{r_0^6} 
 $$
 by the factor $e^{-2\phi(r_0)} \propto r_0^{21/2}$. We obtain
 \be
\label{SBH}
 S \ \propto \ r_0^{9/2}\ .
 \ee
The Hawking temperature is proportional to the so called {\it surface gravity}, i.e. gravitational acceleration at the horizon
(recall the Unruh effect),
 \be
\label{TBH}
T_{\rm Hawking} \propto a_{\rm horizon} \sim \left. \frac {d g_{00}}{dr} \right|_{r = r_0} \propto r_0^{5/2}\ .
 \ee
Combining \eq{SBH} and \eq{TBH}, we obtain $S \propto T^{9/5}$ and hence $\langle E \rangle_T \propto T^{14/5}$.

\end{document}